# Superconducting Hydride $Mg_2RhH_6$ Experimentally Achieved at Lower Pressure

Linjing Wu [a §], Zelong Wang [a §], Guiqi Liu [a §], Jun Zhang [a *], Yanfeng Ge [b], Yuanhao Su [a], Runteng Chen [a], Hongyu Liu [a], Wenmin Li [c], Sijia Zhang [a], Jingcheng Zhu [a], Jianfa Zhao [a], Zheng Deng [a], Shaomin Feng [a], Jing Song [a], Qingqing Liu [a], Xiang Li [d], Haozhe Liu [e], Panpan Kong [a *], Xiancheng Wang [a] and Changqing Jin [a,f *]

[a] *Beijing National Laboratory for Condensed Matter Physics, Institute of Physics, Chinese Academy of Sciences, Beijing,100190, China.*

[b] *State Key Laboratory of Metastable Materials Science and Technology & Hebei Key Laboratory of Microstructural Material Physics, School of Science, Yanshan University, Qinhuangdao, 066004, China.*

[c] *Institute of Quantum Materials and Physics, Henan Academy of Sciences, Zhengzhou 450046, China.*

[d] *Centre for Quantum Physics, Key Laboratory of Advanced Optoelectronic Quantum Architecture and Measurement (MOE), School of Physics, Beijing Institute of Technology, Beijing 100081, China.*

[e] *Center for High Pressure Science & Technology Advanced Research, Beijing, 100094, China.*

[f] *School of Physics, University of Chinese Academy of Sciences, Beijing 100190, China.*

* Email: zhang@iphy.ac.cn

* Email: panpankong@iphy.ac.cn

* Email: jin@iphy.ac.cn

## Abstract

Although tremendous progress has been made in recent years in the field of polyhydride superconductors, the realization of high critical temperature ($T_c$) superconductivity still relies on formidable high pressures. Searching for superconducting hydrides at lower pressures is of particular importance. Here we report the first experimental synthesis of $Mg_2RhH_6$, which exhibits superconductivity under a significantly reduced pressure of 30 GPa. The synthesis of $Mg_2RhH_6$ proceeds via a two-step process: (1) preparing the $Mg_2RhH_5$ precursor, in which hydrogen atoms are stabilized by covalent bonds, and (2) introducing additional hydrogen, which injects electrons into antibonding orbitals above 30 GPa and simultaneously triggers a structural transition from $RhH_5$ square pyramids to $RhH_6$ octahedra. Superconductivity emerges at ~30 GPa with a $T_c$ of 24 K, and the $T_c$ is further enhanced to 29 K upon synthesis at 53 GPa, as evidenced by a sharp drop to zero resistivity and the characteristic suppression of $T_c$ under applied magnetic fields. Our results demonstrate that $Mg_2RhH_6$ is thermodynamically stable above 30 GPa, establishing it as the first superconductor with a $T_c$ of approximately 30 K at a readily accessible pressure. This study pioneers a highly promising pathway for the rational design and discovery of high-temperature superconductors within the phonon-mediated BCS framework.

## Introduction

Hydrogen-rich materials have long been considered potential candidates for high-temperature superconductivity within the phonon-mediated BCS framework, due to the high Debye vibration mode of the lightest element hydrogen. In 1968, Ashcroft proposed that metallic hydrogen might be superconducting with a high superconducting critical temperature ($T_c$) under extremely high-pressure conditions [1]. However, for a long time thereafter, little significant progress was made. In 2004, a novel approach termed "chemical pre-compression" was proposed to achieve hydrogen metallization [2]. By compressing hydrogen-rich compounds, this approach enables the realization of hydrogen metallization and high-temperature superconductivity at relatively low pressures [2]. Subsequently, hydrogen-rich materials, such as $SiH_4$ [3] and $GeH_4$ [4], have become a focus of both theoretical and experimental investigations. A landmark compound, $H_3S$, was initially predicted to be a superconductor with a $T_c$ of approximately 205 K [5]. Subsequently, an experiment quickly confirmed a superconducting transition with a $T_c$ of 203 K in $H_3S$ at ~150 GPa [6]. This finding validated the "chemical pre-compression" theory and accelerated the development of the conventional phonon-mediated hydrogen-based superconductors.

Based on the "chemical pre-compression" theory, numerous binary hydrogen-based superconductors have been designed and have demonstrated high $T_c$s in recent years. Rare earth hydrides (e.g. $LaH_{10}$, $T_c$ = 252 K at 170 GPa [7]; $YH_6$, $T_c$ = 220 K at 183 GPa [8] and $YH_9$, $T_c$ = 243 K at 201 GPa [8]) and alkaline earth metal hydrides (such as $CaH_6$, $T_c$ = 200 K at 160 GPa, 215 K at 172 GPa [9-10]) feature metal atoms with electronegativity significantly lower than that of hydrogen, leading to the formation of hydrogen clathrate structures with predominantly ionic bonding. In contrast, covalent-type hydrogen-rich superconductors like $H_3S$ ($T_c$ = 203 K at 150 GPa) [6], $SbH_4$ ($T_c$ = 116 K at 184 GPa) [11] exhibit primarily covalent bonding under high pressure. Transition metal hydrides, such as $ZrH_6$ ($T_c$ = 71 K at 220 GPa) [12], $HfH_{14}$ ($T_c$ = 83 K at 243 GPa) [13], $NbH_3$ ($T_c$ = 42 K at 187 GPa) [14] and $TaH_3$ ($T_c$ = 30 K at 197 GPa) [15] have been experimentally discovered and stretched hydrogen molecules exist in these materials. Owing to the ultra-high pressures (> 150 GPa) compression, the distance between hydrogen atoms in such materials is relatively small. Taking solid metallic hydrogen as an example, theory predicts that the hydrogen-hydrogen distance ($d_{(H-H)}$) approaches 1 Å at 500 GPa [1].

The core proposition of chemical pre-compression for achieving high-temperature superconductivity involves filling hydrogen's antibonding orbitals with electrons from non-hydrogen elements. This enables density of states of hydrogen to contribute to the Fermi surface. The filling of electrons into antibonding orbitals results in a repulsive interaction in hydride superconductors. Generally, ultra-high pressures are required to enhance the filling of hydrogen's antibonding orbitals, stabilize the structure and achieve a high superconducting $T_c$. Although the $T_c$s of typical hydrides, including $H_3S$ ($d_{(H-H)}$ = 1.5 Å, $T_c$ = 203 K at 150 GPa) [6], $LaH_{10}$ ($d_{(H-H)}$ = 1.1 Å, $T_c$ = 252 K at 170 GPa) [7, 16], $YH_9$ ($d_{(H-H)}$ = 1.1 Å, $T_c$ = 243 K at 201 GPa)[8], $CaH_6$ ($d_{(H-H)}$ = 1.2 Å, $T_c$ = 200 K at 160 GPa, 215 K at 172 GPa)[9-10] have approached room temperature due to the large contribution of hydrogen states at the Fermi surface and strong electron phonon coupling (EPC), the ultra high pressures required hinder the investigations into some physical phenomena and the superconductivity mechanism, such as the Meissner effect. To address this issue, strategies such as chemical doping and covalent stabilization have been explored to reduce the pressures required for achieving superconducting states in hydrogen-rich compounds. For example, the lowest stable pressure for $La_{1-x}Ce_xH_{9-10}$ was reported to be about 100 GPa with a $T_c$ of 176 K [17]. However, these approaches often result in a concurrent decrease in the $T_c$ ($LaH_{10}$, $T_c$ = 252 K at 170 GPa) [7]. Further theoretical studies have predicted a series of multi-component hydrides capable of achieving high $T_c$ under low pressures, such as $LaBH_8$ with a $T_c$ of 156 K at 55 GPa [18], $LaBeH_8$ with a $T_c$ of 98 K at 185 GPa [19]. A later experiment confirmed that $LaBeH_8$ exhibits superconductiviting $T_c$ of 110 K under 80 GPa [20].

To further reduce the pressure required for preparing a new hydride superconductor, we propose that covalent bonding between hydrogen and transition metals with similar electronegativity can stabilize hydrogen atoms at an appropriate distance from each other. Further electron doping fills the antibonding orbitals of both the transition metal and hydrogen, enabling their density of states to contribute to the Fermi surface. Theoretically, this approach can reduce the pressure required for hydrides to obtain superconducting properties. The electronegativity of the noble metals is very close to that of hydrogen, making noble metal hydrides a proper system to validate our proposal [21]. Recently, multiple independent research groups have conducted theoretical investigations on ternary hydrides $Mg_2XH_6$ (where X = noble metal), exhibiting remarkably high $T_c$s ranging from 45 K to 80 K [22-26] and even one prediction reaching as high as 160 K for $Mg_2IrH_6$ [24-27]. Thermodynamic stability and EPC strength calculations show that $Mg_2MH_6$ (M = Rh and Ir) is metastable under ambient pressure and has a high possibility of being synthesized [25, 27]. Usually, the stable transition metal complexes obey the 18 valence electron rule under ambient pressure conditions. $Mg_2MH_6$ (M = Ir and Rh) has 19 valence electrons, while $Mg_2IrH_5$ and $Mg_2RhH_5$ obey the 18-electron rule and therefore should be more thermodynamically stable under ambient pressure conditions. For instance, a series of compounds adhering to the 18-electron rule - namely $Mg_2FeH_6$ [28], $Mg_2CoH_5$ [29], and $Mg_2NiH_4$ [30], have all been successfully synthesized and are stable under ambient conditions. More recently, $Mg_2IrH_5$ was stabilized at ambient pressure [31] and further $Mg_2IrH_7$ near 40 GPa [32], while the synthesis of $Mg_2MH_6$ (M = Rh, Ir) has not yet been achieved.

Motivated by the interest of the $Mg_2Ir/RhH_6$ system, theoretical studies have designed several preparation routes. These routes primarily involve the initial synthesis of either $Mg_2IrH_5$, which has 5/6 hydrogen site occupancy in the $IrH_6$ octahedra [31] or $Mg_2IrH_7$, which has an additional interstitial non-bonding hydrogen atom located at the edge center of the cubic crystal [27], followed by hydrogen insertion or removal. Also, hydrogenating $Mg_2IrH_5$ to form $Mg_2IrH_6$ with the *Amm*2 polymorph was proposed, which is slightly more stable than the $Fm\bar{3}m$ phase at low pressure [27]. Usually, the calculations are performed at zero temperature and $Mg_2Ir/RhH_6$ is predicted to be metastable under ambient pressure [24-25]. The pressurization effect can, to some extent, be equated with lowering the temperature. Therefore, $Mg_2Ir/RhH_6$ might be preserved to room temperature when an appropriate pressure is applied.

$Mg_2IrH_5$ with a *P*4*/nmm* structure and an identical metallic sublattice is exceptionally close to the predicted high-$T_c$ superconductor $Mg_2IrH_6$. Only a single hydrogen atom needs to be added per formula unit to obtain the superconducting phase. Calculations of kinetic barrier indicates that the $Mg_2IrH_5$ (*P*4*/nmm*) system smoothly transitions to the face-centered cubic (fcc) $Mg_2IrH_6$ structure without any energey barrier [31]. Based on the aforementioned ideas, we propose a two-step synthesis method: (1) Synthesis of tetragonal $Mg_2Ir/RhH_5$. Given the close electronegativity between Ir (2.20)/Rh (2.28) and H (2.20), the synthesis of $Mg_2Ir/RhH_5$ can utilize the covalent interaction between Ir/Rh and H to stabilize hydrogen atoms at an appropriate distance (~2.2 Å) [21]. (2) Supplementing hydrogen to prepare the $Mg_2Ir/RhH_6$. Hydrogen injection under pressure-temperature conditions can achieve electron filling into the antibonding orbitals of Ir/Rh and H, potentially enabling their density of states to contribute to the Fermi surface and inducing a superconducting phase transition. Based on sample synthesis, comprehensive structural and electrical property measurements were carried out. The synthesized $Mg_2RhH_6$ sample exhibited a maximum $T_c$ of 29 K, while the synthesis of $Mg_2IrH_6$ is still in progress.

**Experimental part**

Polycrystalline samples of $Mg_2RhH_5$ (*P*4*/nmm*) were prepared by a high-pressure and high-temperature route using a 6 × 1400 T cubic anvil high pressure apparatus. Commercially available crystalline powders of $MgH_2$ (Alfa, 99.9% pure), Rh (Alfa, 99.99% pure) and ammonia borane ($NH_3BH_3$) (Alfa, 97% pure) were used as starting materials. The $MgH_2$ and Rh powders were mixed in a stoichiometric ratio of 2:1 and thoroughly ground, then pressed into a pellet. $NH_3BH_3$ and the mixture were placed in a "sandwich-like configuration" in a sealed Au cylinder and sintered at 900 ℃ under 6 GPa for 30 min to obtain pure polycrystalline samples.

Powder samples of $Mg_2RhH_6$ were prepared by a two-step synthesis route, including first synthesizing $Mg_2RhH_5$ (*P*4*/nmm*) under high pressure conditions and then achieving an additional hydrogen insertion at high pressures by laser heating to 1000-1800 K. $NH_3BH_3$ served as the hydrogen source. The initial pressure, calibrated by the ruby fluorescence method, was set in the range of 30-70 GPa. Different synthesis conditions were attempted, such as pressure, laser power, duration and number of heating cycles. Multiple laser heating spots were distributed across the sample chamber to improve the temperature uniformity for synthesizing $Mg_2RhH_6$ samples. Additionally, dual-sided laser heating was routinely employed to minimize thermal gradients. Temperatures were calibrated spectroradiometrically and monitored in situ via resistance measurements. Reaction completion was confirmed by the stabilization of resistance during iterative heating. The detailed synthesis conditions are summarized in Table SI in the Supporting Information. The nine prepared diamond anvil cell (DAC) samples, labeled as Samples 1-9# were used for transport property measurements. The XRD data were collected based on Sample 10#.

Powder X-ray diffraction (XRD) measurements were carried out on a Rigaku Smart Lab diffractometer with Cu *K*α radiation (λ = 1.54059 Å, 45 kV, 200 mA) in the 2θ range of 4 - 100° with a step size of 0.01°. The diffraction spectra were refined by EXPGUI software [33]. Raman spectra were recorded using a 532 nm laser for excitation. The spectra covered the range of 100–3200 $cm^{-1}$ with a spectral resolution of 1 $cm^{-1}$, employing an integrated Renishaw Raman system equipped with a confocal microscope and a multichannel air-cooled CCD detector. The electronic transport properties were measured by four-probe electrical conductivity methods in a MagLab system with temperatures varying from 300 to 2 K and a magnetic field up to 7 T. The DAC technique was employed for generating high pressure. The beveled diamonds with 300/100 μm culets were chosen and T301 stainless steel was used as the gasket; a mixture of aluminum oxide, sodium chloride, and epoxy resin served as the insulating layer and ammonia borane powder served as the pressure-transmitting medium and hydrogen source. The electric current was set to 0.1 mA during the resistance measurement. The pressure was calibrated by the ruby fluorescence method. The synchrotron radiation XRD experiments under high pressure conditions were performed at the Shanghai Synchrotron Radiation Facility with a radiation wavelength of 0.6199 Å using a symmetric DAC at room temperature. The details of in situ high-pressure experiments can be seen in Refs [11, 34-35].

## Results

**Crystal structure analysis.** As to the synthesized polycrystalline sample of $Mg_2RhH_5$, XRD measurements were carried out with Cu *K*α radiation in the 2θ range from 4° to 100° with a step size of 0.01°. All reflections can be indexed on a tetragonal lattice, as shown in Figure 1(a). The structure refinement was carried out by adopting the initial structural model of $Mg_2CoH_5$ with the space group of *P4/nmm* reported by Zolliker [36], which smoothly converged with $\chi^2$ = 5.24, wRp = 3.73%, Rp = 2.48% and confirmed the high quality of our sample. The lattice constants are *a* = *b* = 4.6164(0) Å, *c* = 6.8233(1) Å and unit cell volume (*V*) of 145.41(0) $Å^3$ when Z=2. Based on the refinement results, a drawing of the main structural unit of $Mg_2RhH_5$ is shown in Figure 2(a-b). Tetragonal $Mg_2RhH_5$ (denoted as $Mg_2RhH_5$-T below) exhibits a distorted $CaF_2$-type metal atom structure and is mainly composed of five-coordinate rhodium ($RhH_5^{4-}$) in an ordered square-pyramidal configuration and $Mg^{2+}$ ions. A $C_{4V}$ symmetry exists in the $RhH_5^{4-}$ unit. The crystallographic data and selected interatomic distances for $Mg_2RhH_5$-T are summarized in Table SII. As expected for a $d^8$ system of this symmetry, the metal-ligand distances toward the base (1.5583(1) Å) are significantly shorter than those toward the apex (1.7256(0) Å), and the bond angle of $\angle H_1$-Rh-$H_1$ is 170.657(1)°. The Rh-H bond lengths are consistent with covalent bonding interactions. The Rh atom is displaced by 0.13 Å from the square pyramidal base toward the apex. The distances of $d_{(H1-H1)}$ and $d_{(H1-H2)}$ within the square pyramidal structure of $Mg_2RhH_5$-T are 2.1964(8) Å and 2.4174(6) Å, respectively.

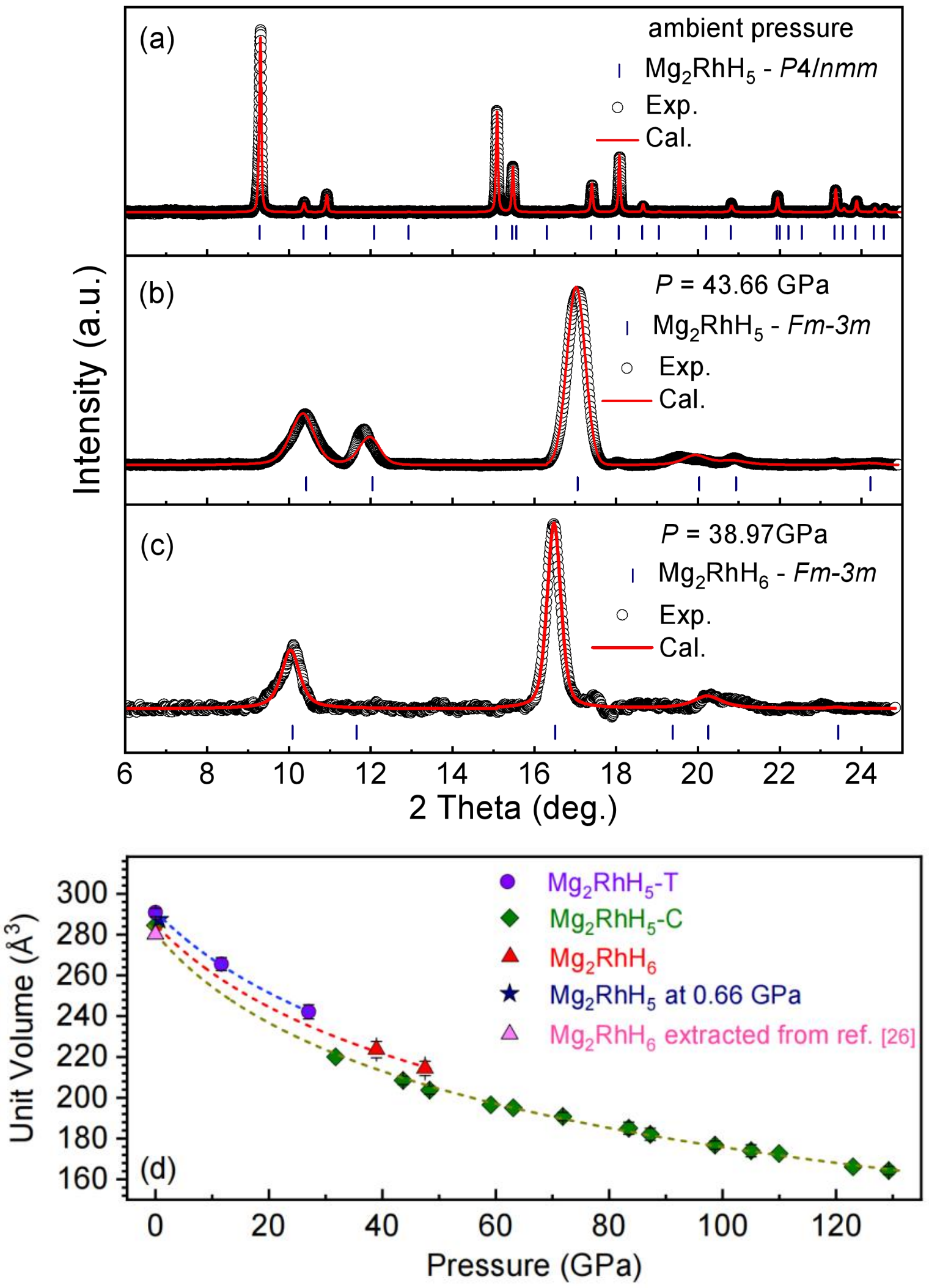

**Figure 1** (a-c) Powder XRD data (represented by circles) and corresponding Rietveld refinement (represented by red lines) for three samples: $Mg_2RhH_5$-T (measured at ambient pressure), $Mg_2RhH_5$ (*Fm*$\bar{3}$*m*, denoted as $Mg_2RhH_5$-C and measured at 43.66 GPa) and $Mg_2RhH_6$ (*Fm*$\bar{3}$*m*, measured at 38.97 GPa). The structural data of $Mg_2RhH_5$-T were collected using Cu *K*α radiation (λ =1.54059 Å) and then converted to synchrotron radiation data (λ = 0.6199 Å) for better comparison with high pressure data. Vertical line bars in the figure mark the positions corresponding to different phases. (d) The pressure and volume relationship for $Mg_2RhH_5$-T, $Mg_2RhH_5$-C and $Mg_2RhH_6$. A phase transition from $Mg_2RhH_5$-T to $Mg_2RhH_5$-C occurred above 30 GPa. $Mg_2RhH_6$ was synthesized from $Mg_2RhH_5$-T using a hydrogen source using DAC equipment at 38.97 GPa after heating to 1000 K. The recovered $Mg_2RhH_5$, obtained by decompressing the $Mg_2RhH_6$ sample synthesized at 38.97 GPa, was measured at 0.66 GPa. Unit-cell volume data are also included. The pressure - volume data are fitted with a second-order Birch-Murnaghan equation of state (EoS) as shown by the dashed lines.

Theoretical calculations predicted that $Mg_2IrH_6$ can be obtained from isostructural $Mg_2IrH_5$ or $Mg_2IrH_7$ by adjusting the hydrogen content [27, 31]. Experimental verification is still lacking. The plausible reason is that this phase may be more metastable rather than thermodynamically stable at ambient pressure and room temperature, and therefore can only be quenched through some non-equilibrium methods. In addition, the tetragonal structure of $Mg_2CoH_5$ can be easily transformed into a disordered cubic structure at ~480 K [36], which is almost isostructural with the theoretically predicted $Mg_2MH_6$ (M= noble metal) [24]. Considering the similar effects of pressure and temperature in the material synthesis, a structural transition under high pressure is also expected. The structure changes of $Mg_2RhH_5$-T under high pressure are investigated. As can be seen from Figure S1, all peaks broaden and systematically shift to higher angles compared with those at ambient pressure as the pressure increases to 27 GPa. An obvious structural transition is observed at 31.76 GPa although the tetragonal structure partially remains as the (111) and (002) peaks of the cubic phase are still asymmetric. When the pressure is further enhanced, the diffraction peaks of (112) and (020) merge and evolve into a cubic structure. These peaks can be well fitted, as shown in Figure 1(b). The crystal lattice is consistently compressed as the pressure is increased to 130 GPa. No new peak appears, which indicates that the crystal structure of $Mg_2RhH_5$ ($Fm\bar{3}m$, denoted as $Mg_2RhH_5$-C below) remains stable in this pressure region.

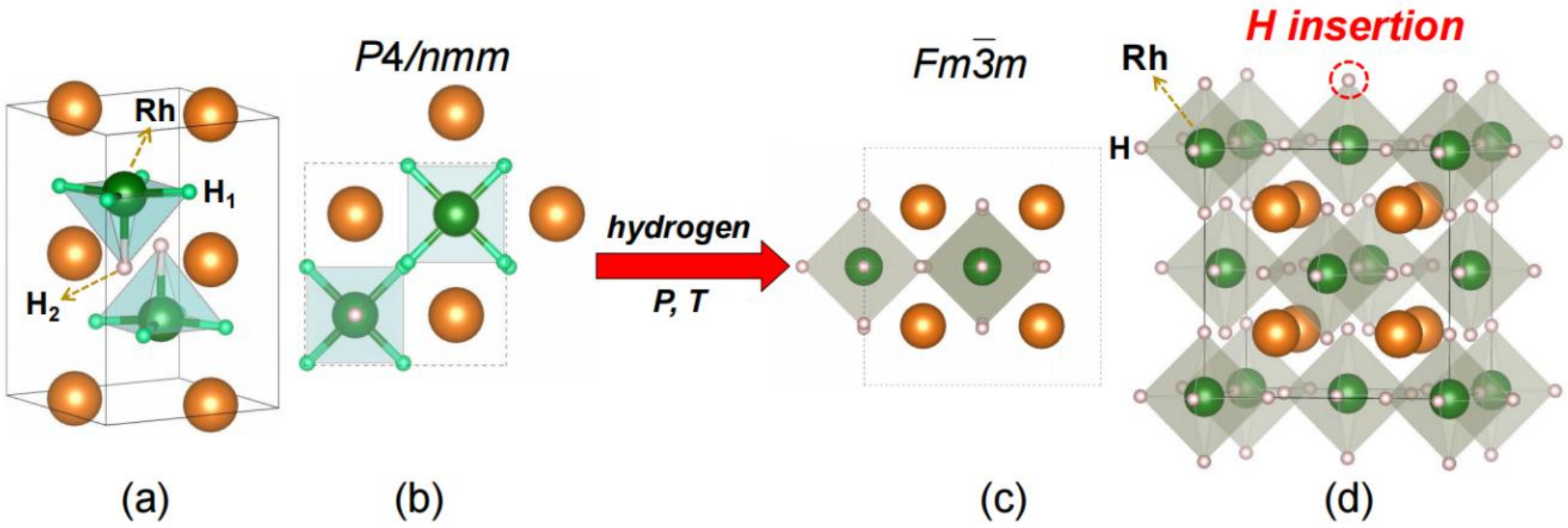

**Figure 2** (a-b) Crystal structure sketches and coordination environment of Rh in $Mg_2RhH_5$ (*P4/nmm*). The Rh ions are in a square-pyramidal configuration, with four $H_1$ atoms in the basal plane and one $H_2$ atom at the apex. (c-d) Crystal structure sketches and coordination environment of Rh in $Mg_2RhH_6$ ($Fm\bar{3}m$), where Rh ions are coordinated by six H atoms forming an ideal octahedral configuration.

All the diffraction data of $Mg_2RhH_5$ were refined, and the typical refinement patterns are shown in Figure S2. Figure 1(d) shows the pressure-volume relationship. At a pressure of 30 GPa, there is an abrupt decrease, indicating a first-order phase transition. That is, tetragonal $Mg_2RhH_5$ with $RhH_5^{4-}$ square-pyramidal configuration transforms to cubic $Mg_2RhH_5$. The $Mg_2RhH_5$-C consists of $RhH_6^{4-}$ octahedra but has a hydrogen occupancy of 5/6 on the octahedral sites. Here, the phase transition is similar to that reported in $Mg_2CoH_5$ [36]. Further compression leads to a smooth reduction in volume. The Birch-Murnaghan equation was used to deduce the bulk modulus $B_0$. The dashed lines are the fitted results obtained with the second-order Birch equation of state (EoS):

$$\mathrm{P(GPa)}=\frac{3}{2}\times \mathrm{B_0}\left[(\mathrm{V_0}/\mathrm{V})^{\frac{7}{3}}-(\mathrm{V_0}/\mathrm{V})^{\frac{5}{3}}\right]\times\left\{1-\left(3-\frac{3}{4}\times \mathrm{B_0'}\right)\times\left[(\mathrm{V_0}/\mathrm{V})^{\frac{2}{3}}-1\right]\right\}$$

where the pressure derivative $B_{0'}$ is set as 4, a fitting to the data up to 27 GPa gives the bulk modulus $B_0$ = 101.19(1) GPa, $V_0$= 291.35(0) $Å^3$ for $Mg_2RhH_5$-T, whereas a fitting to the data with pressure up to 130 GPa gives $B_0$ = 83.50(7) GPa, $V_0$ = 280.66(4) $Å^3$ for $Mg_2RhH_5$-C.

Bond stretching/bending frequencies are sensitive to the local coordination environment, providing a diagnostic probe for distinguishing different molecular complexes. We measured Raman spectra of tetragonal $Mg_2RhH_5$ and by comparison with $Mg_2IrH_5$ [31], observed similar vibrational peaks (Figure S3). However, the stretching mode of Rh–H at ~2200 $cm^{-1}$ severely overlaps with the diamond peak as pressure increased to 15 GPa, making it unresolvable. We thus tracked the Raman peak evolution at 717 $cm^{-1}$ and 1936 $cm^{-1}$ under pressure. As shown in Figure S3(c), an obvious change near 30 GPa corroborates the XRD-detected phase transition from tetragonal to cubic phase. The phase transition is accompanied by the evolution from a $RhH_5^{4-}$ square pyramid to a disordered $RhH_6^{4-}$ octahedron with one hydrogen vacancy.

An attempt was made to prepare $Mg_2RhH_6$ under high pressure conditions where a phase transition was observed. By combining laser heating for hydrogen insertion, it was expected to successfully prepare the target compound. $Mg_2RhH_5$-T and $NH_3BH_3$ (serving as a hydrogen source) were used as the starting materials and the initial pressure was set to 43 GPa. After heating at ~1000 K, the pressure in the DAC for Sample $10^{\#}$ was monitored and calibrated to be 38.97 GPa. The crystal structure was inspected using a synchrotron radiation source with a wavelength of 0.6199 Å. As can be seen from Figure 1(c), the diffraction patterns after heating became very simple, with only two distinct peaks observed. By comparing the diffraction patterns with those of $Mg_2RhH_5$-C at 43.66 GPa in Figure 1(b), the crystal structure can be indexed as the cubic phase of $Mg_2RhH_x$, where x depends on the stoichiometry. After careful refinement, the unit cell volumes at different pressures are shown in Figure 1(d), where the volume curve representing $Mg_2RhH_x$ lies above that of $Mg_2RhH_5$-C. It is well known that determining the hydrogen content in the newly synthesized hydrogen-based superconductors, particularly those prepared by high pressure techniques, remains challenging. Alternatively, as reported in previous studies on metal hydrides, the formation of a metal hydride is typically accompanied by volume expansion relative to the pure metal lattice, owing to hydrogen incorporation. The volume increase upon hydrogenation ranges from 2-3 $Å^3$ [37]. After careful examination, the volume expands by ~2.4(0) $Å^3$ per $Mg_2RhH_x$ unit compared with $Mg_2RhH_5$-C at 38.97 GPa. This result indicates that $Mg_2RhH_6$, the theoretically predicted high-temperature superconductor, may have been synthesized. Equation of state fitting yields a bulk modulus $B_0$ = 96.13(2) GPa, $V_0$= 285.02(0) $Å^3$. Furthermore, Raman spectroscopic measurements, shown in Figure S3, revealed a distinct vibrational mode at around 423 $cm^{-1}$ for the sample containing $Mg_2RhH_5$ and $NH_3BH_3$ after heating at 45.09 GPa. This mode is in good agreement with the $T_{2g}$ vibrational mode of $[IrH_6]^{4-}$ reported in Ref. [32], providing consistent evidence for the formation of $Mg_2RhH_6$.

**Superconductivity property investigations.** The electrical resistance of the $Mg_2RhH_5$-T sample under pressure was first measured, as shown in Figure S4. $Mg_2RhH_5$-T exhibits semiconducting behavior at ambient pressure, which agrees with the calculated results for tetragonal phase $Mg_2IrH_5$ [31]. As pressure increases to ~30 GPa, the electrical resistance sharply decreases to 1.3 Ω at room temperature, and further compression persistently reduces the resistance. The high-pressure XRD patterns in Figure S1 clearly reveal a structural transition from the tetragonal to the cubic phase for $Mg_2RhH_5$-T above ~30 GPa. This transformation is further supported by (i) characteristic changes in the Rh–H vibrational modes, evolving from square-pyramidal $RhH_5^{4-}$ to octahedral $RhH_6^{4-}$ in the Raman spectrum (Figure S3), and (ii) the distinct drop in electrical resistance near ~30 GPa. The sample remains semiconducting up to 54.44 GPa. Therefore, $Mg_2RhH_5$-C with a disordered hydrogen occupation is concluded to be a semiconductor in this pressure range, consistent with the predicted result for cubic $Mg_2IrH_5$ [31].

Hydrogen insertion is expected to induce a metallic phase in $Mg_2RhH_6$. A sample (designated Sample $4^{\#}$) was synthesized at 55 GPa. Electrical resistance measurements were performed to confirm the superconducting transition in $Mg_2RhH_6$. As shown in Figure 3(a), zero resistance is observed in both warming and cooling runs. The $T_c$, determined from the onset of the dR/dT peak, is ~18 K (inset of Figure 3(a)). Notably, residual semiconducting behavior persists in the vicinity of the superconducting transition. The evolution of superconducting transitions under magnetic fields is shown in Figure 3(b), where superconductivity is gradually suppressed to lower temperatures as the field increases to 4 T. This provides solid evidence for superconductivity in $Mg_2RhH_6$.

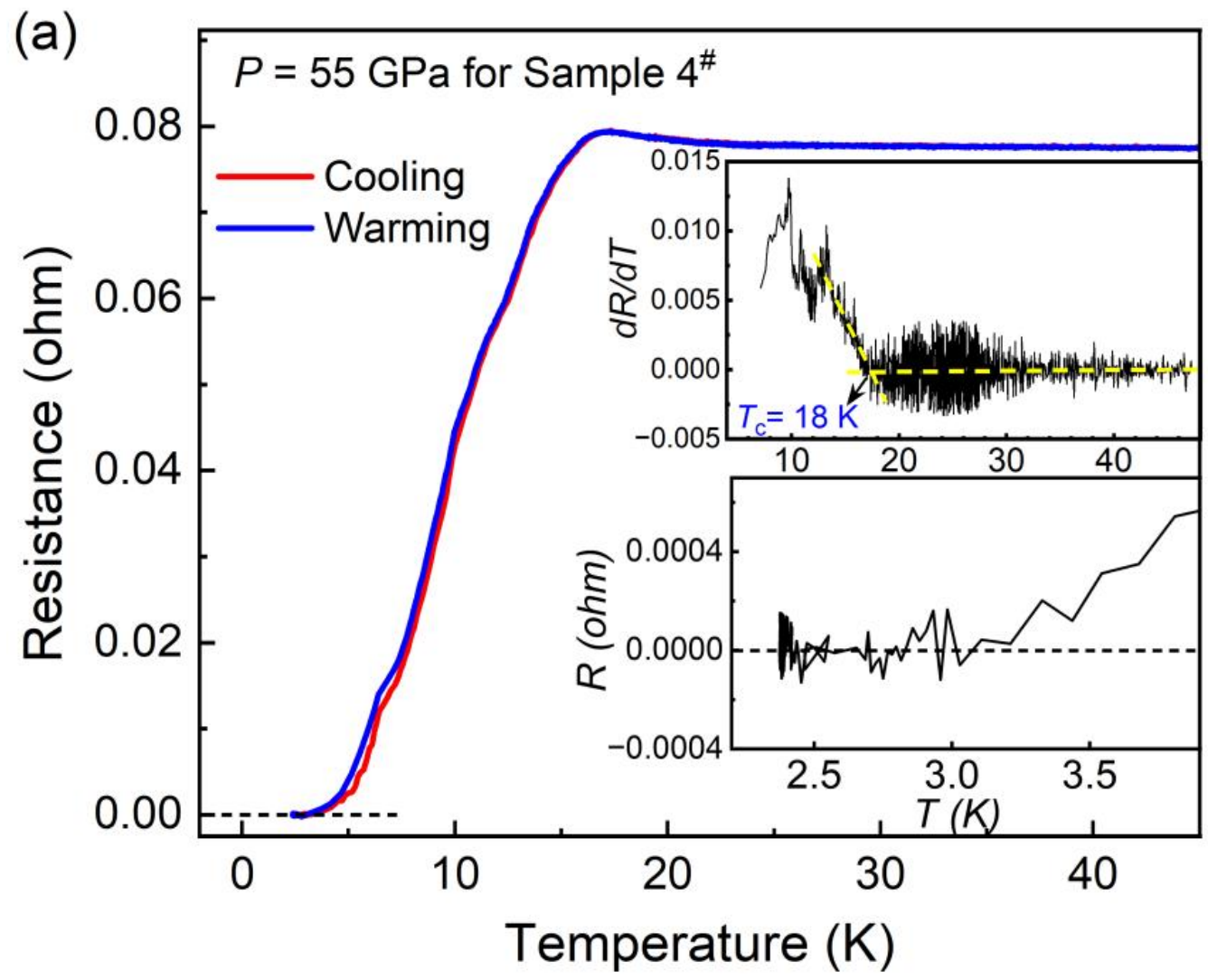

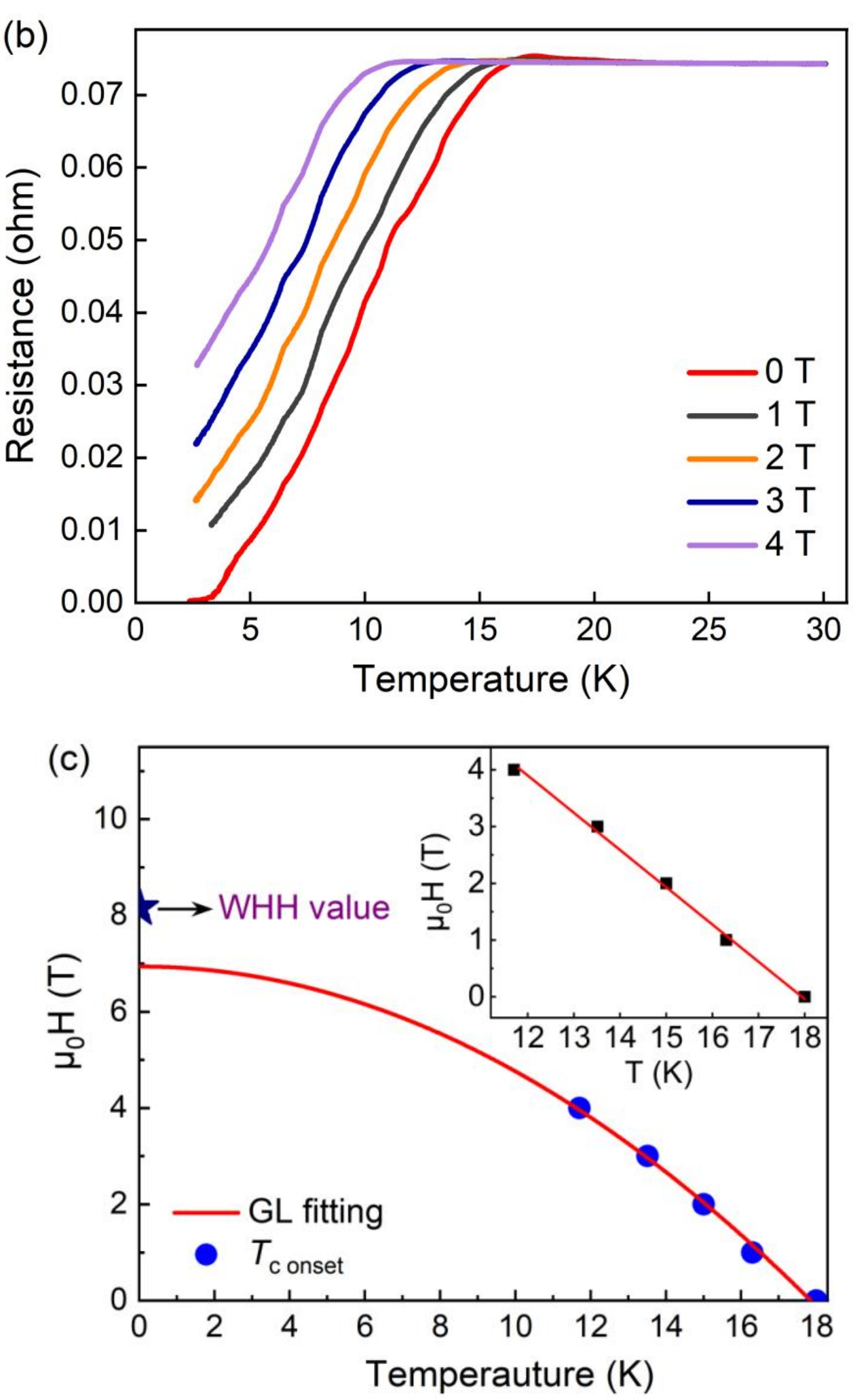

**Figure 3** (a) The temperature-dependent resistance of the Sample 4# measured at 55 GPa during the warming and cooling process. The inset shows the derivative of resistivity (dR/dT) as a function of temperature under zero magnetic field. The superconducting transition temperature can be determined from the turning point of the dR/dT curve. The zero-resistance state is achieved with a residual resistance lower than ±0.0002 Ω, comparable to the instrument noise level. (b) Resistance curves of the Sample 4# measured at 55 GPa under different magnetic fields. (c) The Ginzburg-Landau fitting for the $H_{c2}$(T) is shown by the red solid lines. The star denotes the $H_{c2}(0)$ value calculated via the WHH model. The inset shows the upper critical field $H_{c2}$(T) as a function of temperature.

The superconducting parameters, including upper critical field $H_{c2}(0)$ and Ginzburg-Landau (GL) coherence length ξ, are determined from the measurements on Sample 4#. The $T_{c\ onset}$ under magnetic field ranging from 0 to 4 T is determined using criteria based on the resistance derivative with respect to temperature (dR/dT). The temperature dependence of critical field $H_{c2}$ is shown in the inset of Figure 3(c).

Linear fitting yields a slope of $dH_{c2}/dT|_{Tc}$ = −0.657 T/K for the criteria of $T_{c\ onset}$. Using the Werthamer-Helfand-Hohenberg (WHH) formula, $\mu_0 H_{c2}(T) = -0.69 \times dH_{c2}/dT|_{Tc} \times T_c$, the $H_{c2}(0)$ value was calculated to be 8.16 T. The GL formula, $\mu_0 H_{c2}(T) = \mu_0 H_{c2}(0) \times(1-(T/T_c)^2)$ was also used to estimate the $H_{c2}(0)$ at zero temperature. As shown in Figure 3(c), fitting $\mu_0 H_{c2}(T)$ with the GL formula yields $H_{c2}(0)$ = 6.94 T. Furthermore, the GL coherence length ξ was estimated to be 6.35–6.89 Å using the relation $\mu_0 H_{c2}(0) = \Phi_0/2\pi\xi^2$, where $\Phi_0 = 2.067 \times 10^{-15}$ Web is the magnetic flux quantum and $H_{c2}(0)$ = 6.94–8.16 T from WHH and GL fits. Additionally, high-field electrical transport measurements with magnetic field up to 7 T were performed on Sample 9# (Figure S5). The $\mu_0 H_{c2}(0)$ value obtained from the WHH formula is 9.63 T, which is close to our estimates obtained from Sample 4#.

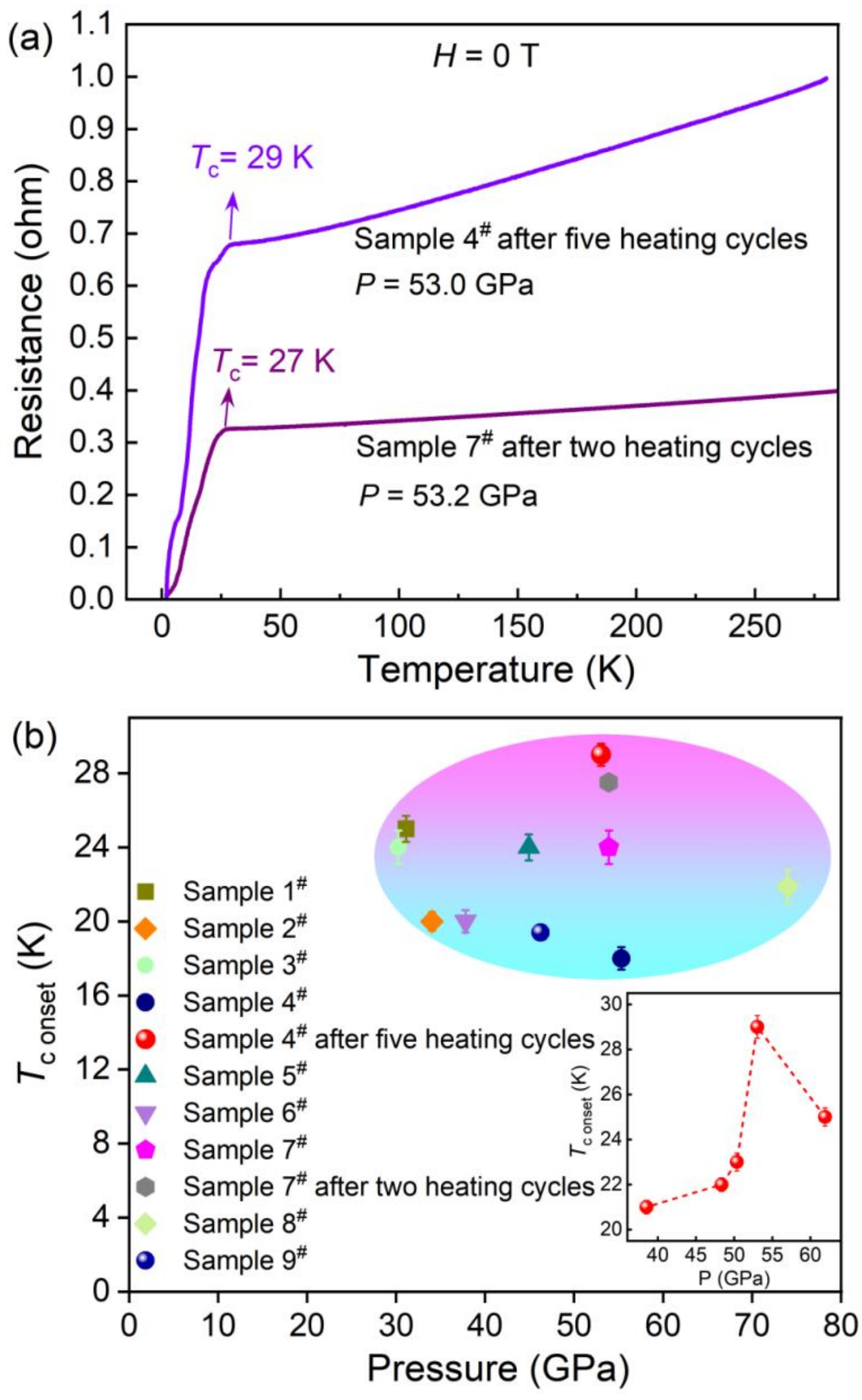

**Figure 4** Superconducting transitions determined by electrical resistance measurements. (a) Temperature dependence of the electrical resistance measured in the cooling process for Sample 4# and Sample 7# within the temperature range of 2-285 K after multiple heating

cycles at high pressure. The arrow indicates the superconducting transition and $T_c$. (b) The pressure dependence of the $T_{c\ onset}$ for $Mg_2RhH_6$ samples (Sample 1-9#) synthesized at 30-74 GPa. The inset shows the $T_c$ evolution upon decompression for the Sample 4# after five heating cycles.

To improve the sample homogeneity of $Mg_2RhH_6$ and enhance the superconducting transition temperature, a series of synthesis experiments were performed, including the preparation of Samples 1-9# at 30-74 GPa, followed by heating to 1000-1800K. Figure 4(a) shows the temperature dependence of the resistance for Sample 4# after five heating cycles, measured during cooling. Metallic behavior is observed in the temperature range from 29 K to room temperature, consistent with the predicted behavior. The resistance quickly drops near 29 K, as indicated by the arrow and approaches zero at low temperatures, suggesting that a superconducting transition occurs in this $Mg_2RhH_6$ sample. A fully metallic normal state and a superconducting transition at 27 K were also observed in sample 7# synthesized at 53.2 GPa after two heating cycles. In contrast, the cubic $Mg_2RhH_5$ still exhibits semiconducting behavior up to 54.44 GPa (Figure S4). This clear distinction further confirms the successful synthesis of the theoretically predicted superconducting $Mg_2RhH_6$ phase.

Figure S6 shows the typical resistance measurements for samples synthesized under various pressure conditions, and details of Samples 1-9# are summarized in Table SI. The drops in electrical resistance are all observed for samples prepared under pressures of 30-74 GPa. For example, Sample 1# exhibits an obvious turning point in electrical resistance at 25 K under 31 GPa, while Sample 8# shows a turning point at 22 K under 74 GPa. Figure 4(b) shows the onset temperatures of the superconducting transition as a function of synthesis pressure (30-74 GPa) for Samples 1-9#, demonstrating intrinsic superconductivity in $Mg_2RhH_6$ with $T_c$ values ranging from 18 to 29 K. The irregular variations in $T_c$ with synthesis pressure are closely associated with hydrogen deficiencies in $Mg_2RhH_{6-x}$ since the introduction of hydrogen atoms serves as electron doping, which dominates the density of states near the Fermi surface and thus $T_c$. Several theoretical studies have predicted $T_c$ values in the range of 45 - 59 K for $Mg_2RhH_6$ [24-26], consistent with our estimated value of ~60 K at 30 - 50 GPa by first-principles calculations (Table SIII), while the experimentally observed maximum $T_c$ in our work is ~29 K. The discrepancy arises from two main factors: (i) variations in the computational methodologies employed in different theoretical works, and (ii) sample inhomogeneities inherent to the high-pressure DAC synthesis technique. In our experiments, the multiple heating cycles result in a gradual increase in $T_c$. We attribute this primarily to the improved sample homogeneity, especially the improvement of inherent hydrogen nonstoichiometry at grain boundaries in the synthesized polycrystalline $Mg_2RhH_6$ samples.

High-pressure AC magnetic susceptibility measurements are highly desirable for observing distinct diamagnetic shielding signals, yet this technique remains challenging within a diamond anvil cell owing to the tiny sample volume and a substantial diamagnetic background signal from the gasket and the cell body. Nevertheless, our extensive transport measurements show excellent reproducibility across multiple runs, and in situ XRD and Raman spectra confirm structural stability. Although the absolute shielding fraction cannot be quantified, these consistent results strongly support the intrinsic superconducting phase of the synthesized $Mg_2RhH_6$ sample.

**Decompression investigations.** Theoretical predictions suggest that the $Mg_2RhH_6$ superconductor is thermodynamically metastable under ambient pressure [24-25]. To investigate this, we conducted decompression experiments on Sample 4#. As shown in Figure S7, at 53.02 GPa the material exhibits a superconducting $T_c$ of 29 K. Upon further compression to 61.91 GPa, $T_c$ decreases to 25 K. During the subsequent decompression, $T_c$ gradually declines—for instance, at 38.51 GPa, it drops to 21 K. When the pressure is reduced to 28.34 GPa, the temperature dependence of the electrical resistance indicates an insulator-to-metal phase transition. As the pressure is further reduced to near-ambient levels, the material becomes completely insulating, consistent with the behavior of $Mg_2IrH_5$ with 5/6 hydrogen occupation. XRD measurements were performed on the $Mg_2RhH_6$ sample synthesized at 38.97 GPa after full decompression. As shown in Figure S8, the pattern collected from the recovered product at 0.66 GPa can be well refined using the cubic structure of $Mg_2RhH_5$ (space group *Fm*-3*m*), with the excellent refinement parameters of $\chi^2 = 0.59$, wRp = 9.92%, and Rp = 7.34%. The refined lattice constant a = 6.5907(1) Å and unit-cell volume V = 286.28(2) Å$^3$ are in good agreement with the experimental values of a = 6.5772(0) Å and V = 284.52(7) Å$^3$ for our independently synthesized cubic $Mg_2RhH_5$, which is a stable phase at ambient pressure as reported for the sister compound $Mg_2IrH_5$ [31-32].

**Discussion**

**Structure stability.** Theoretically, $Mg_2RhH_6$ is predicted to be metastable under ambient pressure, with a formation energy lying ~30 meV above the convex hull [26]. However, our high-pressure structural and electrical transport property measurements confirm that $Mg_2RhH_6$ decomposes into semiconducting $Mg_2RhH_5$ upon decompression below 30 GPa. Based on these results, we suggest that $Mg_2RhH_6$ should be thermodynamically stable only above 30 GPa, which is supported by the negative formation enthalpy and phonon dispersion spectra obtained from calculations above 30 GPa (Figure S9). Furthermore, the experimental results show that $Mg_2RhH_6$ can be synthesized at 30-60 GPa, yet further increasing the synthesis pressure does not enhance its $T_c$. As reported in recent literature [32], $Mg_2IrH_5$ converts directly into $Mg_2IrH_7$ (rather than the theoretically predicted $Mg_2IrH_6$) under high synthesis pressure above 40 GPa with laser heating. It is plausible that $Mg_2RhH_6$ can be synthesized only within a finite high pressure range.

**Superconductivity mechanism.** $Mg_2RhH_6$ adopts a fcc structure, in which the $RhH_6^{4-}$ units form ideal octahedra. These octahedra are entirely isolated from one another, and the shortest hydrogen-hydrogen distance, $d_{(H\text{-}H)}$, which occurs between adjacent $RhH_6^{4-}$ octahedra, measures 2.0601(3) Å at a pressure of 38.97 GPa ($a$ = 6.0697(1) Å, $V$ = 223.61(6) Å$^3$). The distance represents a compression of approximately 7.77 % compared with the ambient pressures (where $d_{(H\text{-}H)}$ = 2.2336(6) Å, $a$ = 6.5809(7) Å, $V_0$ = 285.02(0) Å$^3$, $V_0$ derived from equation-of-state fitting). Concurrently, the Rh-H distance decreases from 1.7110(5) Å to 1.5781(2) Å under compression. Notably, despite this reduction, the $d_{(H\text{-}H)}$ value in $Mg_2RhH_6$ remains significantly larger than those reported in other ionic (e.g. $d_{(H\text{-}H)}$ ~1.2 Å in $CaH_6$ [9]), covalent (e.g. $d_{(H\text{-}H)}$ ~1.5 Å in $H_3S$ [5]), and transition-metal-based hydrogen-rich superconducting materials (e.g. $d_{(H\text{-}H)}$ ~ 0.9 Å in $HfH_{14}$ [13]). Furthermore, it is worth mentioning that the radius of hydrogen ions typically ranges from 1.4 to 2.1 Å [37-38], implying a minimum distance between hydrogen ions of approximately 2.8 Å. This is considerably larger than the observed $d_{(H\text{-}H)}$ value of 2.0601(3) Å in $Mg_2RhH_6$ at 38.97 GPa. Moreover, the Rh-H distance of 1.5781(2) Å at this pressure is characteristic of covalent bonding. These observations together suggest the presence of covalent interactions not only between adjacent hydrogen atoms but also between the central metal Rh and its coordinated hydrogen atoms in $Mg_2RhH_6$. Therefore, the superconductivity observed in $Mg_2RhH_6$ may not be exclusively governed by the density of states of hydrogen at the Fermi surface. As corroborated by theoretical calculations, the density of states near the Fermi level in $Mg_2RhH_6$ is primarily dominated by the $e_g$* antibonding states of the transition metals, which are hybridized with the H-1s and Mg-3s states [24]. This finding aligns with our hypothesis that electrons fill into antibonding orbitals of Rh and H atoms, thereby contributing to the electronic states at the Fermi level. In addition, the phonon linewidth in $Mg_2RhH_6$ is primarily contributed by H modes, which provide at least 2/3 of the integrated coupling constant λ as evidenced by previous theory work [25]. Recently, it was reported that $BiH_2$ exhibits superconductivity with a $T_c$ of approximately 62 K under a pressure of 163 GPa. It was reported that covalent bismuth dominates the electronic states near the Fermi level, and contributes approximately 51% of the total λ [39]. For $Mg_2RhH_6$, the electronic states at the Fermi level are primarily derived from the transition metals, with their contributions significantly surpassing those from hydrogen and magnesium. These electronic states are likely to couple strongly with the high-energy vibrational modes of hydrogen atoms. The novel mechanism suggests the potential for achieving high $T_c$ superconductivity in $Mg_2RhH_6$ based materials, warranting further theoretical exploration to elucidate the underlying physics and optimize material design.

**Conclusion**

In this study, we successfully synthesized the $Mg_2RhH_6$ hydride superconductor for the first time through a two-step route, demonstrating a maximum superconducting $T_c$ of up to 29 K under a pressure of 53 GPa. Our experimental results indicate that the cubic $Mg_2RhH_6$ is stabilized above ca. 30 GPa. The formation of strong covalent bonds between the noble metal (Rh) and hydrogen atoms stabilizes the hydride structure, while the subsequent injection of electrons into the antibonding orbitals induces superconductivity in this material. Further research, such as modifying the covalent bonds by substituting or alloying Rh with other precious metals, offers significant potential to achieve even higher $T_c$ values at pressures closer to ambient conditions.

**Supporting Information:** Experimental details, including material synthesis, high pressure structure and resistance characterizations, decompression experiments and computational details of $Mg_2RhH_5$ and $Mg_2RhH_6$ (PDF).

**Acknowledgements**

This work was supported by the National Key R&D Program of China (Grant No. 2023YFA1406001, 2024YFA1408000, 2023YFA1608902 and 2022YFA1402301) and the National Natural Science Foundation of China (Grant No. 12474097, 12404164). We thank the staff members of the BL15U1 station (https://cstr.cn/31124.02.SSRFBL15U1) and User Experiment Assist System in Shanghai Synchrotron Radiation Facility (SSRF) (https://cstr.cn/31124.02.SSRF).

TOC for Superconducting Hydride $Mg_2RhH_6$ Experimentally Achieved at Lower Pressure

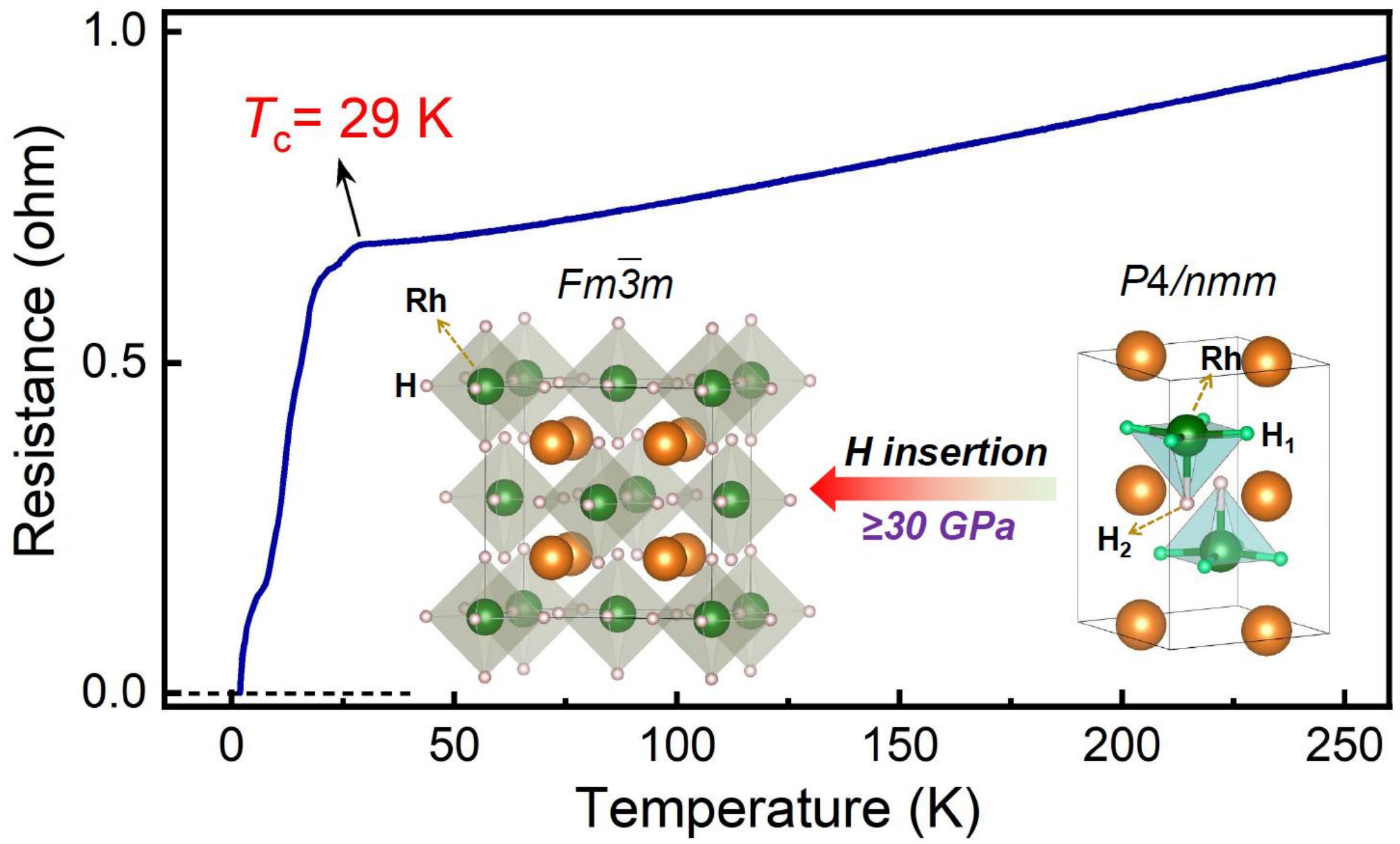